\documentclass[%
 reprint,
superscriptaddress,
 amsmath,amssymb,
 aps,
 prl,
]{revtex4-1}

\usepackage{graphicx}
\usepackage{dcolumn}
\usepackage{bm}

\begin{document}

\preprint{APS/123-QED}
\title{Swelling of doubly magic $^{48}$Ca core in Ca isotopes beyond $N=28$}
\author{M.~Tanaka}
\email{masaomi@phys.kyushu-u.ac.jp}
\affiliation{Department of Physics, Osaka University, Toyonaka, Osaka 560-0043, Japan}
\affiliation{Research Center for Superheavy Elements, Kyushu University, Fukuoka 819-0395, Japan}
\author{M.~Takechi}
\affiliation{Department of Physics, Niigata University, Ikarashi, Niigata 951-2181, Japan}
\author{A.~Homma}
\affiliation{Department of Physics, Niigata University, Ikarashi, Niigata 951-2181, Japan}
\author{M.~Fukuda}
\affiliation{Department of Physics, Osaka University, Toyonaka, Osaka 560-0043, Japan}
\author{D.~Nishimura}
\affiliation{Department of Physics, Tokyo City University, Setagaya, Tokyo 158-8557, Japan}
\author{T.~Suzuki}
\affiliation{Department of Physics, Saitama University, Saitama 338-8570, Japan}
\author{Y.~Tanaka}
\affiliation{Department of Physics, Osaka University, Toyonaka, Osaka 560-0043, Japan}
\author{T.~Moriguchi}
\affiliation{Institute of Physics, University of Tsukuba, Tsukuba, Ibaraki 305-8571, Japan}
\author{D.S.~Ahn}
\affiliation{RIKEN Nishina Center, Wako, Saitama 351-0198, Japan}
\author{A.~Aimaganbetov}
\affiliation{Institute of Nuclear Physics, 050032 Almaty, Kazakhstan}
\affiliation{L.N. Gumilyov Eurasian National University, 010008 Astana, Kazakhstan}
\author{M.~Amano}
\affiliation{Institute of Physics, University of Tsukuba, Tsukuba, Ibaraki 305-8571, Japan}
\author{H.~Arakawa}
\affiliation{Department of Physics, Saitama University, Saitama 338-8570, Japan}
\author{S.~Bagchi}
\affiliation{Astronomy and Physics Department, Saint Mary's University, Halifax, NS B3H 3C3, Canada}
\affiliation{Justus Liebig University, 35392 Giessen, Germany}
\affiliation{GSI Helmholtzzentrum f\"ur Schwerionenforschung, 64291 Darmstadt, Germany}
\author{K.-H.~Behr}
\affiliation{GSI Helmholtzzentrum f\"ur Schwerionenforschung, 64291 Darmstadt, Germany}
\author{N.~Burtebayev}
\affiliation{Institute of Nuclear Physics, 050032 Almaty, Kazakhstan}
\author{K.~Chikaato}
\affiliation{Department of Physics, Niigata University, Ikarashi, Niigata 951-2181, Japan}
\author{H.~Du}
\affiliation{Department of Physics, Osaka University, Toyonaka, Osaka 560-0043, Japan}
\author{S.~Ebata}
\affiliation{School of Environment and Society, Tokyo Institute of Technology, Meguro, Tokyo 152-8551, Japan}
\author{T.~Fujii}
\affiliation{Department of Physics, Saitama University, Saitama 338-8570, Japan}
\author{N.~Fukuda}
\affiliation{RIKEN Nishina Center, Wako, Saitama 351-0198, Japan}
\author{H.~Geissel}
\affiliation{GSI Helmholtzzentrum f\"ur Schwerionenforschung, 64291 Darmstadt, Germany}
\author{T.~Hori}
\affiliation{Department of Physics, Osaka University, Toyonaka, Osaka 560-0043, Japan}
\author{W.~Horiuchi}
\affiliation{Department of Physics, Hokkaido University, Sapporo 060-0810, Japan}
\author{S.~Hoshino}
\affiliation{Department of Physics, Niigata University, Ikarashi, Niigata 951-2181, Japan}
\author{R.~Igosawa}
\affiliation{Department of Physics, Saitama University, Saitama 338-8570, Japan}
\author{A.~Ikeda}
\affiliation{Department of Physics, Niigata University, Ikarashi, Niigata 951-2181, Japan}
\author{N.~Inabe}
\affiliation{RIKEN Nishina Center, Wako, Saitama 351-0198, Japan}
\author{K.~Inomata}
\affiliation{Department of Physics, Saitama University, Saitama 338-8570, Japan}
\author{K.~Itahashi}
\affiliation{RIKEN Nishina Center, Wako, Saitama 351-0198, Japan}
\author{T.~Izumikawa}
\affiliation{Institute for Research Promotion, Niigata University, Niigata 950-8510, Japan}
\author{D.~Kamioka}
\affiliation{Institute of Physics, University of Tsukuba, Tsukuba, Ibaraki 305-8571, Japan}
\author{N.~Kanda}
\affiliation{Department of Physics, Niigata University, Ikarashi, Niigata 951-2181, Japan}
\author{I.~Kato}
\affiliation{Department of Physics, Saitama University, Saitama 338-8570, Japan}
\author{I.~Kenzhina}
\affiliation{Institute of Nuclear Physics, 050032 Almaty, Kazakhstan}
\affiliation{Al-Farabi Kazakh National University, 050040 Almaty, Kazakhstan}
\author{Z.~Korkulu}
\affiliation{RIKEN Nishina Center, Wako, Saitama 351-0198, Japan}
\author{Y.~Kuk}
\affiliation{Institute of Nuclear Physics, 050032 Almaty, Kazakhstan}
\affiliation{L.N. Gumilyov Eurasian National University, 010008 Astana, Kazakhstan}
\author{K.~Kusaka}
\affiliation{RIKEN Nishina Center, Wako, Saitama 351-0198, Japan}
\author{K.~Matsuta}
\affiliation{Department of Physics, Osaka University, Toyonaka, Osaka 560-0043, Japan}
\author{M.~Mihara}
\affiliation{Department of Physics, Osaka University, Toyonaka, Osaka 560-0043, Japan}
\author{E.~Miyata}
\affiliation{Department of Physics, Niigata University, Ikarashi, Niigata 951-2181, Japan}
\author{D.~Nagae}
\affiliation{Research Center for Superheavy Elements, Kyushu University, Fukuoka 819-0395, Japan}
\affiliation{RIKEN Nishina Center, Wako, Saitama 351-0198, Japan}
\author{S.~Nakamura}
\affiliation{Department of Physics, Osaka University, Toyonaka, Osaka 560-0043, Japan}
\author{M.~Nassurlla}
\affiliation{Institute of Nuclear Physics, 050032 Almaty, Kazakhstan}
\affiliation{Al-Farabi Kazakh National University, 050040 Almaty, Kazakhstan}
\author{K.~Nishimuro}
\affiliation{Department of Physics, Saitama University, Saitama 338-8570, Japan}
\author{K.~Nishizuka}
\affiliation{Department of Physics, Niigata University, Ikarashi, Niigata 951-2181, Japan}
\author{K.~Ohnishi}
\affiliation{Department of Physics, Osaka University, Toyonaka, Osaka 560-0043, Japan}
\author{M.~Ohtake}
\affiliation{RIKEN Nishina Center, Wako, Saitama 351-0198, Japan}
\author{T.~Ohtsubo}
\affiliation{Department of Physics, Niigata University, Ikarashi, Niigata 951-2181, Japan}
\author{S.~Omika}
\affiliation{Department of Physics, Saitama University, Saitama 338-8570, Japan}
\author{H.J.~Ong}
\affiliation{Research Center for Nuclear Physics, Osaka University, Ibaraki, Osaka 567-0047, Japan}
\author{A.~Ozawa}
\affiliation{Institute of Physics, University of Tsukuba, Tsukuba, Ibaraki 305-8571, Japan}
\author{A.~Prochazka}
\affiliation{GSI Helmholtzzentrum f\"ur Schwerionenforschung, 64291 Darmstadt, Germany}
\author{H.~Sakurai}
\affiliation{RIKEN Nishina Center, Wako, Saitama 351-0198, Japan}
\affiliation{Department of Physics, University of Tokyo, Bunkyo-ku, Tokyo 113-0033, Japan}
\author{C.~Scheidenberger}
\affiliation{GSI Helmholtzzentrum f\"ur Schwerionenforschung, 64291 Darmstadt, Germany}
\author{Y.~Shimizu}
\affiliation{RIKEN Nishina Center, Wako, Saitama 351-0198, Japan}
\author{T.~Sugihara}
\affiliation{Department of Physics, Osaka University, Toyonaka, Osaka 560-0043, Japan}
\author{T.~Sumikama}
\affiliation{RIKEN Nishina Center, Wako, Saitama 351-0198, Japan}
\author{H.~Suzuki}
\affiliation{RIKEN Nishina Center, Wako, Saitama 351-0198, Japan}
\author{S.~Suzuki}
\affiliation{Institute of Physics, University of Tsukuba, Tsukuba, Ibaraki 305-8571, Japan}
\author{H.~Takeda}
\affiliation{RIKEN Nishina Center, Wako, Saitama 351-0198, Japan}
\author{Y.K.~Tanaka}
\affiliation{GSI Helmholtzzentrum f\"ur Schwerionenforschung, 64291 Darmstadt, Germany}
\author{I.~Tanihata}
\affiliation{Research Center for Nuclear Physics, Osaka University, Ibaraki, Osaka 567-0047, Japan}
\affiliation{School of Physics and Nuclear Energy Engineering, Beihang University, 100191 Beijing, China}
\author{T.~Wada}
\affiliation{Department of Physics, Niigata University, Ikarashi, Niigata 951-2181, Japan}
\author{K.~Wakayama}
\affiliation{Department of Physics, Saitama University, Saitama 338-8570, Japan}
\author{S.~Yagi}
\affiliation{Department of Physics, Osaka University, Toyonaka, Osaka 560-0043, Japan}
\author{T.~Yamaguchi}
\affiliation{Department of Physics, Saitama University, Saitama 338-8570, Japan}
\author{R.~Yanagihara}
\affiliation{Department of Physics, Osaka University, Toyonaka, Osaka 560-0043, Japan}
\author{Y.~Yanagisawa}
\affiliation{RIKEN Nishina Center, Wako, Saitama 351-0198, Japan}
\author{K.~Yoshida}
\affiliation{RIKEN Nishina Center, Wako, Saitama 351-0198, Japan}
\author{T.K.~Zholdybayev}
\affiliation{Institute of Nuclear Physics, 050032 Almaty, Kazakhstan}
\affiliation{Al-Farabi Kazakh National University, 050040 Almaty, Kazakhstan}
\date{\today}
\begin{abstract}
Interaction cross sections for $^{42\textrm{--}51}$Ca on a carbon target at 280~MeV/nucleon have been measured for the first time. The neutron number dependence of derived root-mean-square matter radii shows a significant increase beyond the neutron magic number $N=28$. Furthermore, this enhancement of matter radii is much larger than that of the previously measured charge radii, indicating a novel growth in neutron skin thickness. A simple examination based on the Fermi-type distribution, and the Mean-Field calculations point out that this anomalous enhancement of the nuclear size beyond $N=28$ results from an enlargement of the core by a sudden increase in the surface diffuseness of the neutron density distribution, which implies the swelling of the bare $^{48}$Ca core in Ca isotopes beyond $N=28$.
\begin{description}
\item[PACS numbers]25.60.Dz
\end{description}
\end{abstract}

\pacs{Valid PACS appear here}
\maketitle

Systematic studies of nuclear radii along the isotopic chain have so far elucidated changes in the nuclear structure such as the emergence of a halo as well as the development of neutron skin and nuclear deformation~\cite{TA85a,SU95,OZ01,TA12,TA14}. Nuclear charge radii, which represent charge spreads in these nuclei, also give complemental information on the size of the nucleus. It has been revealed that the trend of charge radii along the isotopic chain shows a sudden increase, which is often called a ``kink,'' just after the magic number~\cite{AN13}. In particular, the neutron magic number $N=28$ has received considerable attention. Recently, unexpectedly large charge radii were observed in neutron-rich Ca isotopes beyond $N=28$~\cite{GA16}. This sudden growth in charge radii from $^{48}$Ca ($N=28$) to $^{52}$Ca represents a challenging problem; it has not been quantitatively explained by any theoretical calculations other than the Hartree--Fock--Bogolyubov calculation with the Fayans energy density functional~\cite{MI19}. This anomalous phenomenon observed in Ca isotopes is stimulating further studies of nuclear charge radii in a wide mass region~\cite{MI19,MI16,HA18,TR18,GO19}.

In contrast, information on the evolution of the size of the neutron density distribution has not been obtained across $N=28$. For example, nucleon density distributions $\rho_\textrm{m}(r)$ or point-neutron density distributions $\rho_\textrm{n}(r)$ for Ca isotopes have been deduced only for stable nuclei, $^{40,42,44,48}$Ca, through the hadron elastic scattering~\cite{FR68,AL76,AL77,CH77,IG79,RA81,AL82,BO84,MC86,GI92,ZE18}.

The experimental data for root-mean-square (RMS) radii of $\rho_\textrm{m}(r)$ or $\rho_\textrm{n}(r)$ for Ca isotopes beyond $N=28$ are helpful to understand this anomalous phenomenon. Therefore, we have performed measurements of interaction cross sections for Ca isotopes across $N=28$. Interaction cross section $\sigma_\textrm{I}$ or reaction cross section $\sigma_\textrm{R}$ is an observable sensitive to the RMS radius of nucleon density distribution $\langle r^2 \rangle^{1/2}_\textrm{m}$ (hereinafter referred to as the ``matter radius''). The $\sigma_\textrm{I}$ measurements have played a crucial role in the discovery of halo and skin structures~\cite{TA85a,SU95,OZ01}. Recent $\sigma_\textrm{I}$ measurements also revealed the strong deformation of neutron-rich Ne and Mg isotopes and the existence of deformed-halo nuclei~\cite{TA12,TA14}.

In this letter, we report the first $\sigma_\textrm{I}$ measurements for $^{42\textrm{--}51}$Ca across the neutron magic number $N=28$, performed at the RIKEN Radioactive Isotope Beam Factory (RIBF). The matter radii were derived from the $\sigma_\textrm{I}$ data by using the Glauber model calculation. Moreover, by combining these with existing charge radii, the neutron skin thicknesses $\Delta r_\textrm{np}$ were derived. A dramatic enhancement of matter radii beyond $N=28$ was observed; it is similar to the growth in the charge radii of Ca isotopes but is much greater in magnitude.

The experiment was conducted at the RIBF, operated by the RIKEN Nishina Center, RIKEN, and the Center for Nuclear Study, University of Tokyo. Secondary beams of $^{42\textrm{--}51}$Ca were produced with a 345~MeV/nucleon $^{238}$U primary beam bombarding a rotating beryllium production target installed at the F0 focal plane of the BigRIPS superconducting fragment separator~\cite{KU12}. The secondary beams produced were roughly purified at the first stage of the BigRIPS separator, which corresponds to the beam line between the F0 and F3 focal planes. 

After purification of the secondary beams at the first stage of the BigRIPS separator, $\sigma_\textrm{I}$ was measured by the transmission method~\cite{TA13} between the F3 and F7 focal planes. In the transmission method, $\sigma_\textrm{I}$ is derived through the equation, $\sigma_\textrm{I}=-(1/N_\textrm{t})\ln(\Gamma/\Gamma_\textrm{0})$, where $N_\textrm{t}$ is the number of target nuclei per unit area, and $\Gamma$ and $\Gamma_0$ are the non-reaction rates with and without the reaction target, respectively. For the achromatic focus on the F7 focal plane, a wedge-shaped natural carbon target (the wedge angle is 9.61~mrad) was set at the F5 momentum-dispersive focal plane as a reaction target. The target thickness is 1.803(3)~g/cm$^2$ at the central point. In the $\sigma_\textrm{I}$ measurement with such a wedge-shaped target, $\sigma_\textrm{I}$ can be obtained from their values at each position, $\sigma_\textrm{I}(x)$, by weighting with the distribution of incident particles on the target $N_\textrm{in}(x)$, where $x$ is the momentum-dispersive (horizontal) direction perpendicular to the beam axis. The profile of target thickness $t(x)$ was measured with an accuracy of 0.15\% or better. The mean energy in the reaction target, $E_\textrm{ave}$, at the weighted mean position of $N_\textrm{in}(x)$ is 280~MeV/nucleon.

The non-reaction rates were derived by counting the incoming particles before the reaction target and the non-reacting particles after the reaction target. For this purpose, incoming and outgoing particles were identified in an event-by-event mode between the F3 and F5 focal planes, and between the F5 and F7 focal planes, respectively, by combining the magnetic rigidity~($B\rho$), time-of-flight~(TOF), and energy loss~($\Delta{E}$) in the same manner as explained in Refs.~\cite{TA12,TA14}. These quantities were measured by three kinds of detectors: plastic scintillation counters (PL) at F3, F5, and F7; parallel plate avalanche counters (PPAC) at F3; and multisampling ionization chambers (MUSIC) at F3, F5, and F7.

\begin{figure}[t]
\resizebox{0.5\textwidth}{!}{\includegraphics{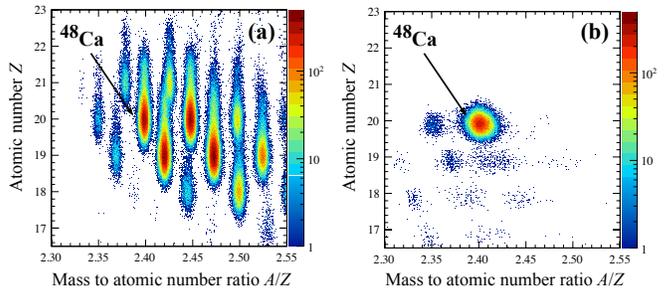}}
\caption{Particle-identification plots of (a) a cocktail beam including $^{48}$Ca before the reaction target and (b) outgoing particles after the reaction target with the selection of incoming $^{48}$Ca. The appropriate constraint in the beam emittance is adopted in both plots.}
\label{pid1}
\end{figure}

As an example, Fig.~\ref{pid1}(a) shows a particle-identification (PID) plot for the beam before the reaction target in the case of $^{48}$Ca. With the 1--10~pnA primary beam, the typical total beam intensity of the cocktail beams on the reaction target was $3\times10^3$~pps, which corresponds to $1.8\times10^2$, $4.5\times10^2$, $2.8\times10^2$, and $3.0\times10^1$~pps for $^{48\textrm{--}51}$Ca. The typical PID resolution is $6.2\sigma$ so that the objective nuclides are well separated from neighboring nuclides. From this PID plot, the number of incident $^{48}$Ca particles was counted. The number of non-reacting particles after the reaction target was counted in Fig.~\ref{pid1}(b), which is the PID plot for the beam after the reaction target with the selection of incoming $^{48}$Ca. The position, angle, and momentum information obtained from upstream detectors were constrained to assure the full transmission for non-reacting particles after the reaction target. This constraint was optimized for each objective nuclide.

\begin{table}
\caption{Measured interaction cross sections $\sigma_\textrm{I}$ for $^{42\textrm{--}51}$Ca on a C target at 280~MeV/nucleon.}
\begin{ruledtabular}
\begin{tabular}{cccc}
 & $\sigma_\mathrm{I}$ & & $\sigma_\mathrm{I}$ \\
Nuclide & (mb) & Nuclide & (mb)\\ \hline
$^{42}$Ca & 1463(14) & $^{47}$Ca & 1509(17) \\
$^{43}$Ca & 1476(12) & $^{48}$Ca & 1498(18) \\
$^{44}$Ca & 1503(13) & $^{49}$Ca & 1561(12) \\
$^{45}$Ca & 1481(9)  & $^{50}$Ca & 1615(15) \\
$^{46}$Ca & 1505(11) & $^{51}$Ca & 1650(42) \\
\end{tabular}
\end{ruledtabular}
\label{tab1}
\end{table}

The present $\sigma_\textrm{I}$ for $^{42\textrm{--}51}$Ca on $^{12}$C at 280~MeV/nucleon are summarized in Table~\ref{tab1}. The statistical error is typically less than $\sim$1.0\%, whereas the total systematic error, caused mainly by the validity of the adopted event selection, is at most 0.4\%.

To discuss the present results along with existing RMS charge radii $\langle r^2 \rangle^{1/2}_\textrm{ch}$~\cite{GA16}, $\langle r^2 \rangle^{1/2}_\textrm{m}$ were deduced using the Glauber model calculation (called the MOL[FM] calculation in Ref.~\cite{TA09}). As $\sigma_\textrm{I}$ is almost the same as $\sigma_\textrm{R}$ above $\sim100$~MeV/nucleon, the Glauber model calculation can be adopted to derive $\langle r^2 \rangle^{1/2}_\textrm{m}$ from $\sigma_\textrm{I}$~\cite{TA13}. The MOL[FM] calculation can reproduce the experimental $\sigma_\textrm{R}$ for $^{12}$C on $^{9}$Be, $^{12}$C, and $^{27}$Al targets, whose $\rho_\textrm{m}(r)$ are well known, with deviations much less than 1\% at $200\textrm{--}300$~MeV/nucleon~\cite{TA09}. Other Glauber models~\cite{HO07,TR16} give almost the same cross section values in this energy region. The $\rho_\textrm{m}(r)$ introduced in Ref.~\cite{TA09} was used as the nucleon-density profile of the target nucleus, $^{12}$C, which provides a good reproduction of the energy dependence of $\sigma_\textrm{R}$ for $^{12}$C on $^{12}$C~\cite{TA09}. 

For the projectile nuclei, $\rho_\textrm{m}(r)$ was assumed to be the two-parameter Fermi-type (2pF) function:
\begin{equation}
\rho_\textrm{m}(r)=\frac{\rho_0}{1+\exp\left[\left(r-C_\textrm{m}\right)/a_\textrm{m}\right]},
\label{EqFermi}
\end{equation}
where $\rho_0$ is the density constant, $C_\textrm{m}$ is the half-density radius, and $a_\textrm{m}$ is the diffuseness. Although this function has three parameters, the known quantities are the measured $\sigma_\textrm{I}$ and the mass number $A=\int \rho_\textrm{m}(r)d^3r$. Therefore, an additional restriction is required. Based on the characteristics of nuclear matter, the saturation density is almost constant in any nuclide. From this point of view, the central density $\rho_\textrm{m}(0)=\rho_0/[1+\exp(-C_\textrm{m}/a_\textrm{m})]$ is constrained to $0.176\textrm{ fm}^{-3}$. Based on the above model function, $\langle r^2 \rangle^{1/2}_\textrm{m}$ was obtained so as to reproduce the measured $\sigma_\textrm{I}$. The value of $\rho_\textrm{m}(0)$ was determined as the weighted mean of available data for experimental $\rho_\textrm{m}(0)$ of $^{40,42,44,48}$Ca measured through the elastic scattering with hadronic probes~\cite{FR68, AL76, AL77, CH77, RA81, AL82, BO84}. The corresponding standard deviation around the weighted mean value, $\Delta\rho_\textrm{m}(0)$, results in a systematic error of 0.020~fm in $\langle r^2 \rangle^{1/2}_\textrm{m}$. As this systematic error is smaller than the typical statistical errors of the present results, it does not affect the following discussion. Even if we fix $C_\textrm{m}$ or $a_\textrm{m}$ to a typical value (for instance, $C_\textrm{m}=1.2A^{1/3}$~fm or $a_\textrm{m}=0.5$~fm) as another additional restriction instead of the constraint on $\rho_\textrm{m}(0)$, the obtained $\langle r^2 \rangle^{1/2}_\textrm{m}$ agrees well with the value obtained from $\rho_\textrm{m}(0)=0.176\textrm{ fm}^{-3}$ within the error bars. Therefore, the adopted assumption likewise does not influence the following discussion.

For simplicity in the following discussion, the existing charge radii~\cite{GA16} were converted to the RMS radii of the point-proton density distribution $\langle r^2 \rangle^{1/2}_\textrm{p}$. This procedure was performed using $\langle r^2 \rangle_\textrm{p}=\langle r^2 \rangle_\textrm{ch}-R^2_\textrm{p}-(N/Z)R^2_\textrm{n}-3\hbar^2/(4m^2_\textrm{p}c^2)$, where $R_\textrm{p}$ and $R_\textrm{n}$ are the RMS charge radii of the proton and neutron, respectively [$R_\textrm{p}=0.8751(61)$~fm~\cite{PA16}, $R_\textrm{n}^2 = -0.1149(24)$~fm$^2$~\cite{KO97}], and $3\hbar^2/(4m_\textrm{p}^2c^2)$ represents the Darwin--Foldy correction term~\cite{FR97}.

\begin{figure}[t]
\resizebox{0.5\textwidth}{!}{\includegraphics{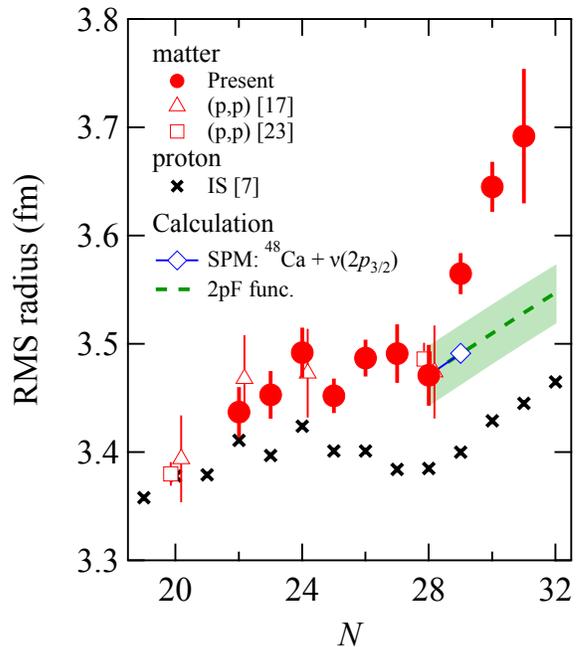}}
\caption{Neutron number dependence of the root-mean-square radii of nucleon density distributions $\langle r^2 \rangle^{1/2}_\textrm{m}$ and those of point-proton density distributions $\langle r^2 \rangle^{1/2}_\textrm{p}$~\cite{GA16} for Ca isotopes. The present results are represented by closed circles, whereas the existing experimental results of $\langle r^2 \rangle^{1/2}_\textrm{m}$ deduced from proton elastic scattering (p,p)~\cite{IG79,ZE18} are plotted as open triangles~\cite{IG79} and open squares~\cite{ZE18}, respectively. The experimental results of $\langle r^2 \rangle^{1/2}_\textrm{p}$ measured using the isotope-shift (IS) method~\cite{GA16} are represented by crosses. The examinations by the single-particle model (SPM) and the two-parameter Fermi-type (2pF) function are also indicated by the open diamond connected by the solid line and the dashed line with the green band, respectively.}
\label{Fig2}
\end{figure}
The derived $\langle r^2 \rangle^{1/2}_\textrm{m}$ values are plotted as filled circles in Fig.~\ref{Fig2}. The present results for $^{42,44,48}$Ca are consistent with the existing results obtained through the proton elastic scattering, represented by open triangles~\cite{IG79} and open squares~\cite{ZE18}. Figure~\ref{Fig2} also shows existing $\langle r^2 \rangle^{1/2}_\textrm{p}$ of $^{39\textrm{--}52}$Ca~\cite{GA16}, represented by crosses. Comparing $\langle r^2 \rangle^{1/2}_\textrm{m}$ with $\langle r^2 \rangle^{1/2}_\textrm{p}$, common features of a sudden increase beyond $N=28$ as well as of a staggering around $N=24$ are seen. Note that the mass excesses of Ca isotopes also show similar features~\cite{WA17}. In contrast, the significant difference between $\langle r^2 \rangle^{1/2}_\textrm{m}$ and $\langle r^2 \rangle^{1/2}_\textrm{p}$ is much larger enhancement beyond $N=28$ in $\langle r^2 \rangle^{1/2}_\textrm{m}$; for example, the difference between the values of $\langle r^2 \rangle^{1/2}_\textrm{p}$ for $^{48}$Ca and $^{50}$Ca is 0.044(4)~fm, whereas the difference between the values of $\langle r^2 \rangle^{1/2}_\textrm{m}$ is 0.17(4)~fm.

First, to understand the mechanism of this enhancement, the effect of the halo-like structure was examined. A halo nucleus, such as $^{11}$Li, whose matter radius is greatly enhanced~\cite{TA85a}, usually has an \emph{s}- or \emph{p}-wave loosely bound valence nucleon. In the Ca isotopic chain beyond $N=28$, the valence neutron configuration changes from the $1f_{7/2}$ to the $2p_{3/2}$ orbital, which is supported by the experimental results of magnetic moments~\cite{GA15}. Therefore, the wave function of the valence neutron may spread spatially. For example, considering the case of $^{49}$Ca, $\langle r^2 \rangle^{1/2}_\textrm{m}$ was calculated based on a single-particle model (SPM)~\cite{TA10a} with the combination of a bare doubly magic $^{48}$Ca core and a $2p_{3/2}$ valence neutron. The wave function of the valence neutron was calculated using the Woods--Saxon potential. The potential depth was tuned to reproduce the experimental one-neutron separation energy $S_\textrm{n}$ in the same manner as explained in Ref.~\cite{TA10a}. 

The calculated $\langle r^2 \rangle^{1/2}_\textrm{m}$ of $^{49}$Ca is represented in Fig.~\ref{Fig2} by the open diamond connected by the solid line. This calculation does not explain the remarkable increase from $^{48}$Ca to $^{49}$Ca. This is because $S_\textrm{n}$ values of $^{49\textrm{--}51}$Ca are not so small ($S_\textrm{n}=4.8\textrm{--}6.4$~MeV)~\cite{WA17}. That is, their valence neutrons are strongly bound compared with the typical halo nucleus ($S_\textrm{n}<1\textrm{ MeV}$). To reproduce the experimental enhancement within this SPM, the valence neutron should have an RMS radius of 6.7(9)~fm, which is comparable to that of two-valence neutrons in $^{11}$Li~\cite{TA92}. Such an incredible spread can be achieved with an unrealistic value of $S_\textrm{n}\sim0.0$~MeV. Hence, this enhancement cannot be attributed to solely the effect of the excess neutrons outside the bare $^{48}$Ca. To explain the experimental result of $^{49}$Ca, based on this SPM with the valence-neutron wave function to reproduce the experimental $S_\textrm{n}$, a significant core enlargement of 0.08(4)~fm is needed.

Another possible reason for the enhancement of $\langle r^2 \rangle^{1/2}_\textrm{m}$ could be deformation of the nuclei~\cite{TA12,TA14}. However, according to the experimental electric quadrupole moments of $^{49,51}$Ca~\cite{GA15}, their quadrupole deformations are nearly 0 owing to the proton magicity of $Z=20$. 

We also studied the origin of the enhancement of $\langle r^2 \rangle^{1/2}_\textrm{m}$ beyond $N=28$ by using a simple 2pF function of Eq.~(\ref{EqFermi}). Here, we assume that the excess neutrons contribute to form a neutron skin entirely outside the bare $^{48}$Ca by regarding $C_\textrm{m}$ as the only variable against $A$. Based on the characteristics of density distributions of the stable nuclei, $\rho_\textrm{m}(0)$ and $a_\textrm{m}$ are almost constant in any nuclide. From this point of view, $\langle r^2 \rangle^{1/2}_\textrm{m}$ of the 2pF function for $^{49\textrm{--}51}$Ca were calculated under the assumptions of $\rho_\textrm{m}(0)=0.176$~fm$^{-3}$ and $a_\textrm{m}=0.49$~fm. The value of $\rho_\textrm{m}(0)$ is the same as already mentioned, whereas that of $a_\textrm{m}$ was determined to reproduce the present $\langle r^2 \rangle^{1/2}_\textrm{m}$ of $^{48}$Ca. Under these conditions, $C_\textrm{m}$ was determined from the mass number, $A=\int \rho_\textrm{m}(r)d^3r$, for the respective nuclides.

The results calculated using this model are represented by the dashed line and the green band in Fig.~\ref{Fig2}. Likewise this examination also cannot reproduce the $N$ dependence of experimental $\langle r^2 \rangle^{1/2}_\textrm{m}$ beyond $N=28$. Although this picture seems to result in the maximum increase in $\langle r^2 \rangle^{1/2}_\textrm{m}$ while retaining the bare $^{48}$Ca, the increase in the experimental values surprisingly exceeds that given by this picture. Note that the experimental enhancement of $\langle r^2 \rangle^{1/2}_\textrm{m}$ from $^{48}$Ca to $^{49}$Ca corresponds to the five-neutron excess outside $^{48}$Ca under this model.

Examinations with both the SPM and the 2pF function indicate that excess neutrons stimulate the bare $^{48}$Ca to swell in Ca isotopes beyond $N=28$. In other words, the doubly magic core, $^{48}$Ca, seems not to be retained in their nuclides. Considering the examination using the 2pF function, there are two extreme possibilities to explain the trend of the aforementioned experimental results: the decrease in $\rho_\textrm{m}(0)$ or the increase in $a_\textrm{m}$ along the isotopic chain beyond $N=28$. For example, in order to reproduce the experimental enhancement of $\langle r^2 \rangle^{1/2}_\textrm{m}$ from $^{48}$Ca to $^{51}$Ca, $\rho_\textrm{m}(0)$ must change from 0.176~fm$^{-3}$ to 0.149~fm$^{-3}$, while $a_\textrm{m}$ must change from 0.49~fm to 0.60~fm.

\begin{figure}[t]
\resizebox{0.5\textwidth}{!}{\includegraphics{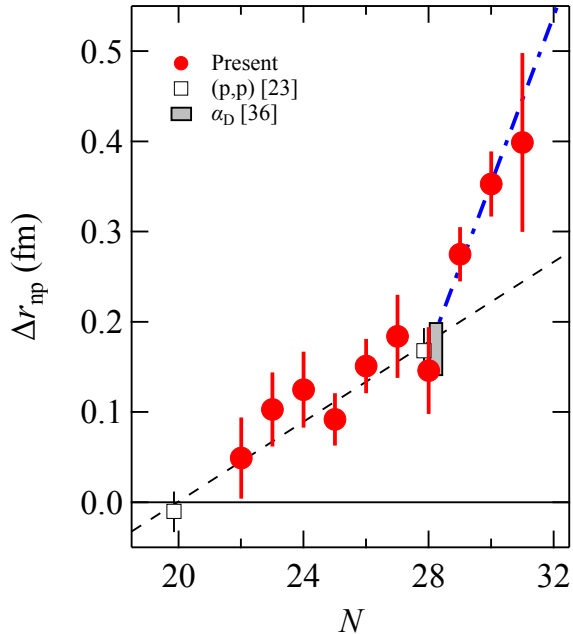}}
\caption{Neutron number dependence of neutron skin thickness $\Delta r_\textrm{np}$ for Ca isotopes. The present results are represented by filled circles, and the existing experimental results deduced from electric dipole polarizability $\alpha_\textrm{D}$~\cite{BI17} and those from proton elastic scattering (p,p)~\cite{ZE18} are represented by the rectangle and by open squares, respectively. The dashed and dashed-and-dotted lines are the results of the chi-square fitting to the data below and above  $N = 28$.}
\label{Fig3}
\end{figure}

The neutron skin thickness $\Delta r_\textrm{np}$ is a sensitive observable for further study to shed light on the difference between protons and neutrons. The present results and existing $\langle r^2 \rangle^{1/2}_\textrm{p}$ data enable us to derive $\Delta r_\textrm{np}$ of $^{42\textrm{--}51}$Ca from the following equation:
\begin{equation}
	\Delta r_\textrm{np} = \sqrt{ \cfrac{A\langle r^2 \rangle_\textrm{m}-Z\langle r^2 \rangle_\textrm{p}}{N}} - \langle r^2 \rangle^{1/2}_\textrm{p},
\end{equation} 
where the first term on the right-hand side represents the RMS radius of the point-neutron density distribution $\langle r^2 \rangle^{1/2}_\textrm{n}$. The derived $\Delta r_\textrm{np}$ values are shown in Fig.~\ref{Fig3}. The present $\Delta r_\textrm{np}$ result for $^{48}$Ca is in good agreement with the existing experimental values deduced not only from the measurement of electric dipole polarizability $\alpha_\textrm{D}$~\cite{BI17} (shaded rectangle) but also from the proton elastic scattering~\cite{ZE18} (open squares). Furthermore, it is revealed that $\Delta r_\textrm{np}$ of Ca isotopes exhibits the same striking increase beyond $N=28$ as that for $\langle r^2 \rangle^{1/2}_\textrm{m}$, indicating that the enhancement of  $\langle r^2 \rangle^{1/2}_\textrm{n}$ beyond $N=28$ is much stronger than that of $\langle r^2 \rangle^{1/2}_\textrm{p}$~\cite{GA16}. Accordingly, the enhancement of $\langle r^2 \rangle^{1/2}_\textrm{p}$ beyond the magic number seems to simply reflect the effect of the novel growth in neutron skin via an attractive force between protons and neutrons~\cite{GO13}.

Note that the Mean-Field (MF) calculations also predict a sizable kink at the magic numbers, including $N=28$ of Ca isotopes, in the trend of $\Delta r_\textrm{np}$ as well as $\langle r^2 \rangle^{1/2}_\textrm{n}$ and $\langle r^2 \rangle^{1/2}_\textrm{p}$~\cite{HOR17,WA10,WA14}. According to these calculations, such a kink structure results from the sudden increase in the surface diffuseness in $\rho_\textrm{n}(r)$; this supports one of the possibilities suggested in the abovementioned discussion on the present $\langle r^2 \rangle^{1/2}_\textrm{m}$ using the 2pF function. Therefore, the doubly magic core, $^{48}$Ca, in Ca isotopes beyond $N=28$ swells with the increase in the surface diffuseness in $\rho_\textrm{n}(r)$. The present results experimentally indicate that the change in $\rho_\textrm{n}(r)$ plays the main role in the enhancement of the nuclear size including charge radii beyond the magic numbers.

This phenomenon of the enhancement of the nuclear size is newly recognized as a mechanism apart from the halo and nuclear deformation mechanisms. We point out that a signature similar to the one found in the present results may also exist in the $\langle r^2 \rangle^{1/2}_\textrm{m}$ change of isotopic chains around oxygen at $N=14$~\cite{KA01,KA11}, whose mechanism is still controversial. A similar suggestion is also proposed by the theory~\cite{NA19}. On the other hand, the sudden increase in charge radii is not observed at $N=2$, $8$, or $20$, which are magic numbers independent of the spin-orbit force. Therefore, such a novel growth in the nuclear size seems to be a universal phenomenon in the magic or semi-magic numbers where a $j_>=l+1/2$ orbital is just closed.

In summary, we performed $\sigma_\textrm{I}$ measurements for $^{42\textrm{--}51}$Ca on a natural carbon target at 280~MeV/nucleon. From these data, $\langle r^2 \rangle^{1/2}_\textrm{m}$ values were derived using the Glauber model calculation. Although the derived values show a neutron number dependence similar to that in $\langle r^2 \rangle^{1/2}_\textrm{p}$, the enhancement of $\langle r^2 \rangle^{1/2}_\textrm{m}$ beyond $N=28$ is much stronger. Furthermore, we obtained $\Delta r_\textrm{np}$ by combining the present data with the existing $\langle r^2 \rangle^{1/2}_\textrm{p}$ values. The values deduced for $\Delta r_\textrm{np}$ also increase dramatically beyond $N=28$. This is because the enhancement of $\langle r^2 \rangle^{1/2}_\textrm{n}$ beyond $N=28$ is radically greater than that of $\langle r^2 \rangle^{1/2}_\textrm{p}$. From the point of view of not only the examination using the 2pF function but also the MF calculations, the kink in the trend of nuclear size at the magic number is thought to result from the sudden increase in the surface diffuseness of $\rho_\textrm{n}(r)$.

\begin{acknowledgments}
We would like to express our gratitude to the accelerator staff of the RIKEN Nishina Center for providing the intense $^{238}$U beam. The present work was supported in part by the Grant-in-Aid for JSPS KAKENHI Grant No.~JP24244024 and No.~JP16H03905 and for JSPS Research Fellow Grant No.~JP15J01446.
\end{acknowledgments}

\nocite{*}


\begin{thebibliography}{9999}
\bibitem{TA85a} I. Tanihata \emph{et al.}, Phys. Rev. Lett. \textbf{55}, 2676 (1985).
\bibitem{SU95} T. Suzuki \emph{et al.}, Phys. Rev. Lett. \textbf{75}, 3241 (1995).
\bibitem{OZ01} A. Ozawa \emph{et al.}, Nucl. Phys. A \textbf{693}, 32--62 (2001).
\bibitem{TA12} M. Takechi \emph{et al.}, Phys. Lett. B \textbf{707}, 357--361 (2012).
\bibitem{TA14} M. Takechi \emph{et al.}, Phys. Rev. C \textbf{90}, 061305(R) (2014).
\bibitem{AN13} I. Angeli and K. P. Marinova, At. Data Nucl. Data Tables \textbf{99}, 69--95 (2013).
\bibitem{GA16} R. F. Garcia Ruiz \emph{et al.}, Nature Phys. \textbf{12}, 594--598 (2016).
\bibitem{MI19} A. J. Miller \emph{et al.}, Nature Phys. \textbf{15}, 432--436 (2019).
\bibitem{MI16} K. Minamisono \emph{et al.}, Phys. Rev. Lett. \textbf{117}, 252501 (2016).
\bibitem{HA18} M. Hammen \emph{et al.}, Phys. Rev. Lett. \textbf{121}, 102501 (2018).
\bibitem{TR18} D. T. Tran \emph{et al.}, Nature Comm. \textbf{9}, 1594 (2018).
\bibitem{GO19} C. Gorges \emph{et al.}, Phys. Rev. Lett. \textbf{122}, 192502 (2019).
\bibitem{FR68} R. F. Frosch \emph{et al.}, Phys. Rev. \textbf{174}, 1380--1399 (1968).
\bibitem{AL76} G. D. Alkhazov \emph{et al.}, Nucl. Phys. A \textbf{274}, 443--462 (1976).
\bibitem{AL77} G. D. Alkhazov \emph{et al.}, Nucl. Phys. A \textbf{280}, 365--376 (1977).
\bibitem{CH77} A. Chaumeaux \emph{et al.}, Phys. Lett. B \textbf{72}, 33--36 (1977).
\bibitem{IG79} G. Igo \emph{et al.}, Phys. Lett. B \textbf{81}, 151--155 (1979).
\bibitem{RA81} L. Ray \emph{et al.}, Phys. Rev. C \textbf{23}, 828--837 (1981).
\bibitem{AL82} G. D. Alkhazov \emph{et al.}, Nucl. Phys. A \textbf{381}, 430--444 (1982).
\bibitem{BO84} K. G. Boyer \emph{et al.}, Phys. Rev. C \textbf{29}, 182--194 (1984).
\bibitem{MC86} R. H. McCamis \emph{et al.}, Phys. Rev. C \textbf{33}, 1624--1633 (1986).
\bibitem{GI92} W. R. Gibbs and J.-P. Dedonder, Phys. Rev. C \textbf{46}, 1825--1833 (1992).
\bibitem{ZE18} J. Zenihiro \emph{et al.}, arXiv:1810.11796 (2018).
\bibitem{KU12} T. Kubo \emph{et al.}, Prog. Theor. Exp. Phys. \textbf{2012}, 03C003 (2012).
\bibitem{TA13} I. Tanihata \emph{et al.}, Prog. Part. Nucl. Phys. \textbf{68}, 215--313 (2013). 
\bibitem{TA09} M. Takechi \emph{et al.}, Phys. Rev. C \textbf{79}, 061601(R) (2009).
\bibitem{HO07} W. Horiuchi \emph{et al.}, Phys. Rev. C \textbf{75}, 044607 (2007).
\bibitem{TR16} D. T. Tran \emph{et al.}, Phys. Rev. C \textbf{94}, 064604 (2016).
\bibitem{PA16} C. Patrignani \emph{et al.}, Chin. Phys. C \textbf{40}, 100001 (2016).
\bibitem{KO97} S. Kopecky \emph{et al.}, Phys. Rev. C \textbf{56}, 2229--2237 (1997).
\bibitem{FR97} J. L. Friar \emph{et al.}, Phys. Rev. A \textbf{56}, 4579--4586 (1997).
\bibitem{WA17} M. Wang \emph{et al.}, Chin. Phys. C \textbf{41}, 030003 (2017).
\bibitem{GA15} R. F. Garcia Ruiz \emph{et al.}, Phys. Rev. C \textbf{91}, 041304(R) (2015).
\bibitem{TA10a} K. Tanaka \emph{et al.}, Phys. Rev. C \textbf{82}, 044309 (2010).
\bibitem{TA92} I. Tanihata \emph{et al.}, Phys. Lett. B \textbf{287}, 307--311 (1992).
\bibitem{BI17} J. Birkhan \emph{et al.}, Phys. Rev. Lett. \textbf{118}, 252501 (2017).
\bibitem{GO13} P.M. Goddard \emph{et al.}, Phys. Rev. Lett. \textbf{110}, 032503 (2013).
\bibitem{WA10} M. Warda \emph{et al.}, Phys. Rev. C \textbf{81}, 054309 (2010).
\bibitem{WA14} M. Warda \emph{et al.}, Phys. Rev. C \textbf{89}, 064302 (2014).
\bibitem{HOR17} W. Horiuchi \emph{et al.}, Phys. Rev. C \textbf{96}, 035804 (2017).
\bibitem{KA01} R. Kanungo \emph{et al.}, Phys. Lett. B \textbf{512}, 261--267 (2001).
\bibitem{KA11} R. Kanungo \emph{et al.}, Phys. Rev. C \textbf{84}, 061304(R) (2011).
\bibitem{NA19} H. Nakada, submitted to Phys. Rev. C.

\end{thebibliography}

\end{document}